\documentclass[prd,showpacs,preprintnumbers,superscriptaddress,floatfix,twocolumn,amsmath,amssymb]{revtex4}
\usepackage{epsfig}
\usepackage{graphicx}
\usepackage{dcolumn}
\usepackage{bm}
\newcommand{\be}{\begin{equation}}

\newcommand{\ee}{\end{equation}}
\newcommand{\bea}{\begin{eqnarray}}
\newcommand{\eea}{\end{eqnarray}}

\begin{document}


\title{Kernel method for nonlinear Granger causality}
\date{\today}
\author{Daniele~Marinazzo}
\author{Mario~Pellicoro}
\author{Sebastiano~Stramaglia}
\affiliation{Dipartimento  Interateneo di Fisica, Universit\`a di
Bari, Italy,} \affiliation{Istituto Nazionale di Fisica Nucleare,
Sezione di Bari, Italy}\affiliation{TIRES-Center of Innovative
Technologies for Signal Detection and Processing, Universit\`a di
Bari,  Italy,}

\date{\today}

\begin{abstract}
Important information on the structure of complex systems,
consisting of more than one component, can be obtained by measuring
to which extent the individual components exchange information among
each other. Such knowledge is needed to reach a deeper comprehension
of phenomena ranging from turbulent fluids to neural networks, as
well as complex physiological signals. The linear Granger approach,
to detect cause-effect relationships between time series, has
emerged in recent years as a leading statistical technique to
accomplish this task. Here we generalize Granger causality to the
nonlinear case using the theory of reproducing kernel Hilbert
spaces. Our method performs linear Granger causality in the feature
space of suitable kernel functions, assuming arbitrary degree of
nonlinearity. We develop a new strategy to cope with the problem of
overfitting, based on the geometry of reproducing kernel Hilbert
spaces. Applications to coupled chaotic maps and  physiological data
sets are presented.

\end{abstract}

\maketitle Experiments in many fields of science provide time series
of simultaneously recorded variables. The analysis of the
synchronization between  time series \cite{synch} is an important
tool to study communications between different components of a
complex systems. In many systems, however, it is important not only
to detect synchronized states, but also to identify cause-effect
(drive-response) relationships between components
\cite{hla,lungarella}. Information-theoretic approaches to causality
are based on the estimation of entropy and mutual information
\cite{schreiber,wiese,frenzel}, an hard numerical problem when
conditioning with respect to a large number of variables is to be
done. Another major approach to analyze causality between two time
series has been proposed by Granger \cite{granger}: if the
prediction error of the first time series is reduced by including
measurements from the second one in the linear regression model,
then the second time series is said to have a causal influence on
the first one. This linear frame for measuring causality has been
widely applied in many fields, including rheochaos \cite{reo}
neurophysiology \cite{ding}, economy \cite{economy}, and climatology
\cite{triacca}. The importance of Granger causality is the
suggestion to use prediction (and  tools from {\it learning theory}
\cite{vapnik-book-1998} in particular) to measure the amount of
information exchanged by two (sub)systems; it is worth mentioning
that also a measure of self-organization, rooted on optimal
predictors, has been recently proposed \cite{shalizi}. Some attempts
to extend Granger causality to the nonlinear case have been recently
proposed \cite{nl}. The main problem of all approaches is detection
of false causalities \cite{palus}, which may arise due to
over-fitting of the learning scheme.

The purpose of this work is to present a novel approach which
measures Granger causality of time series, assuming arbitrary degree
of nonlinearity, while controlling overfitting, and thus avoiding
the problem of false causalities. To this aim we exploit the
properties of kernel machines, the state-of-the-art in learning
models \cite{shawe}.

We start describing the connection between Granger causality and
information-theoretic approaches like the transfer entropy $T_E$ in
\cite{schreiber}. Let $\{\xi_n\}_{n=1,.,N+m}$ be a time series that
may be approximated by a stationary Markov process of order $m$,
i.e.
$p(\xi_n|\xi_{n-1},\ldots,\xi_{n-m})=p(\xi_n|\xi_{n-1},\ldots,\xi_{n-m-1})$.
We will use the shorthand notation
$X_i=(\xi_{i},\ldots,\xi_{i+m-1})^\top$ and $x_i=\xi_{i+m}$, for
$i=1,\ldots,N$, and treat these quantities as $N$ realizations of
the stochastic variables $X$ and $x$. The minimizer of the risk
functional, $ R \left[ f \right] = \int dX dx \left( x - f(X)
\right)^2 p(X, x)$, represents the best estimate of $x$, given X,
and corresponds \cite{papoulis} to the regression function $f^*(X) =
\int dx p(x | X )x$. Now, let $\{\eta_n\}_{n=1,.,N+m}$ be another
time series of simultaneously acquired quantities, and denote
$Y_i=(\eta_{i},\ldots,\eta_{i+m-1})^\top$. The best estimate of $x$,
given $X$ and $Y$, is now: $g^*(X,Y)=\int dx  p(x | X,Y )x$. If the
generalized Markov property holds, i.e. \be \label{genmarkov}p(x |
X,Y )=p(x | X ),\ee then $f^*(X)=g^*(X,Y)$ and the knowledge of $Y$
does not improve the prediction of $x$. $T_E$ \cite{schreiber} is a
measure of the violation of \eqref{genmarkov}: it follows that
Granger causality implies non-zero transfer entropy.

Due to the finiteness of $N$, the risk functional cannot be
evaluated; we consider the empirical risk $\mbox{\emph{ER}}\left[ f
\right]=\sum_{i=1}^N \left(x_i - f(X_i)\right)^2,$ and the search
for the minimum of \mbox{\emph{ER}} is constrained in a suitable
functional space, called hypothesis space; the simplest choice  is
the space of all linear functions, corresponding to linear
regression. In the following we propose a geometrical description of
linear Granger causality. For each $\alpha\in \{1,\ldots,m\}$, the
samples of the $\alpha$-th component of $X$ form a  vector ${\bf
u}_\alpha \in \Re^N$; without loss of generality we assume that each
${\bf u}_\alpha$ has zero mean and that
$\bf{x}$=$(x_1,\ldots,x_N)^\top$ is normalized and zero mean. We
denote $\tilde{x_i}$ the value of the linear regression of $x$
versus $X$, evaluated at $X_i$. The vector
$\tilde{\bf{x}}$=$(\tilde{x_1},\ldots,\tilde{x_N})^\top$ can be
obtained as follows. Let $H\subseteq \Re^N$ be the span of
$\textbf{u}_1,\ldots,\textbf{u}_m$; then $\tilde{\bf{x}}$ is the
projection of $\bf{x}$ on $H$. In other words, calling $P$ the
projector on the space $H$, we have $\tilde{\bf{x}}=P \bf{x}$.
Moreover, the prediction error, given $X$, is
$\epsilon_x=||\bf{x}-\tilde{\bf{x}}||^2=1-\tilde{\bf{x}}^\top
\tilde{\bf{x}}$. Calling $\bf{X}$ the $m\times N$ matrix having
vectors $\bf{u}_\alpha$ as rows, $H$ coincides with the range of the
$N\times N$ matrix $\textbf{K}=\textbf{X}^\top \textbf{X}$.

Using both $X$ and $Y$, the values of the linear regression form the
vector $\tilde{\bf{x}}^\prime=P^\prime\bf{x}$, $P^\prime$ being the
projector on the space $H^\prime\subseteq \Re^N$, spanned by the
$\textbf{u}_1,\ldots,\textbf{u}_m$ and the components of $Y$
$\textbf{v}_1,\ldots,\textbf{v}_m$ (assumed to be zero mean).
$H^\prime$ is the range of the matrix
$\textbf{K}^\prime=\textbf{Z}^\top \textbf{Z}$, where $\bf{Z}$ is
the $2m\times N$ matrix with vectors $\bf{u}_\alpha$ and
$\bf{v}_\alpha$ as rows.  The prediction error is now
$\epsilon_{xy}=||\bf{x}-\tilde{\bf{x}}^\prime||^2=1-\tilde{\bf{x}}^{\prime
\top} \tilde{\bf{x}}^\prime.$ We now note that $H\subseteq
H^\prime$, hence $H^\prime=H\oplus H^\perp$. The last formula shows
geometrically the enlargement of the hypothesis space, due to the
inclusion of the $Y$ variables. Calling $P^\perp$ the projector on
$H^\perp$, we have: $\epsilon_{xy}=\epsilon_x - ||P^\perp
\textbf{x}||^2$, and the linear Granger causality index reads: \be
\delta\left( Y\to X\right)={\epsilon_{x}-\epsilon_{xy}\over
\epsilon_{x}}={||P^\perp \bf{x}||^2\over 1-\tilde{\bf{x}}^\top
\tilde{\bf{x}}}.\label{delta1} \ee Linear Granger causality is
usually assessed according to well known test statistics, see e.g.
\cite{hla}. Instead of assessing the presence  (or not) of causality
by means of a single statistical test, and in view of the non-linear
extension, we introduce a causality index which by construction is
not affected by over-fitting. We observe that $H^\perp$ is the range
of the matrix
$\tilde{\bf{K}}=\bf{K}^\prime-P\bf{K}^\prime-\bf{K}^{\prime}P+P\bf{K}^{\prime}P.$
Hence the {\it natural} choice of the orthonormal basis in $H^\perp$
is the set of the eigenvectors, with non vanishing eigenvalue, of
$\tilde{\bf{K}}$. Calling $\bf{t}_1,\ldots,\bf{t}_{m}$ these
eigenvectors, we have: $||P^\perp {\bf x}||^2=\sum_{i=1}^m r_i^2$,
where $r_i$ is the Pearson's correlation coefficient of $\bf{x}$ and
$\bf{t}_i$. To avoid false causalities, we first evaluate,  by
Student's t test, the probability $\pi_i$ that $r_i$ is due to
chance, assuming $\bf{x}$ and $\bf{t}_i$ normal. Since we are
dealing with multiple comparison, we use the Bonferroni correction
to select the eigenvectors $\bf{t}_{i^\prime}$, correlated with
$\bf{x}$, with expected fraction of false positive equal to 0.05.
Then we calculate a new causality index by summing only over the
$\{r_{i^\prime}\}$ which pass the Bonferroni test, thus obtaining
what we call {\it filtered} linear Granger causality index: \be
\delta_F\left( Y\to X\right)={\sum_{i^\prime} r_{i^\prime}^2\over
1-\tilde{\bf{x}}^\top \tilde{\bf{x}}}.\label{delta1}\ee Exchanging
the roles of the two time series, we may evaluate the causality
index in the opposite direction $\delta_F\left( X\to Y\right)$.

The formulation of linear Granger causality, above described, allows
an efficient generalization to the nonlinear case using methods of
the theory of Reproducing Kernel Hilbert Spaces (RKHS) \cite{shawe}.
Let us first deal with the problem of predicting $x$ using the
knowledge of $X$. Given a kernel function $k$, with spectral
representation $k(X,X^\prime)=\sum_a \lambda_a \Psi_a (X)\Psi_a
(X^\prime)$, we consider $H$, the range of the $N\times N$ Gram
matrix $\bf{K}$ with elements $K_{ij}=k(X_i,X_j)$. As in the linear
case, we calculate $\tilde{\bf{x}}$, the projection of $\bf{x}$ onto
$H$. Due to the spectral representation of $k$, $\tilde{\bf{x}}$
coincides with the linear regression of $\bf{x}$ in the feature
space spanned by  $\sqrt{\lambda_a}\Psi_a$, the eigenfunctions of
$k$; the regression is nonlinear in the original variables. We
remark that $H$ corresponds to the functional space where well known
methods, like Support Vector Machines and Kernel Ridge Regression,
search for the regressor \cite{shawe}.

While using both $X$ and $Y$ to predict $x$, we append $X$ and $Y$
variables to construct the $Z$ variable with samples $Z_i=(X_i
Y_i)^\top$; then we evaluate the Gram matrix $\bf{K^\prime}$ with
elements $K^\prime_{ij}=k(Z_i,Z_j)$. The regression values now form
vector $\tilde{\bf{x}}^\prime$ equal to the projection of $\bf{x}$
on $H^\prime$, the range of $\bf{K^\prime}$. In this work we
consider two choices of the kernel (see the discussion in
\cite{ancona}): the inhomogeneous polynomial (IP) of integer order
$p$: $ k_p(X,X^\prime)=\left(1+X^\top X^\prime\right)^p, $ and the
Gaussian: $
k_\sigma(X,X^\prime)=\exp{\left(-{\left(X-X^\prime\right)^\top
\left(X-X^\prime\right)\over 2\sigma^2}\right)},$ whose complexity
depends on the scale parameter $\sigma$.

First we consider the IP kernel. In this case the eigenfunctions
$\Psi_a$ are all the monomials, in the input variables, up to the
$p-th$ degree. In this case $H\subseteq H^\prime$, and we can
proceed as in the linear case, decomposing $H^\prime=H\oplus
H^\perp$ and calculating
$\tilde{\bf{K}}=\bf{K}^\prime-P\bf{K}^\prime-\bf{K}^{\prime}P+P\bf{K}^{\prime}P.$
Along the same lines as those described in the linear case, we may
construct the filtered Granger causality taking into account only
the eigenvectors of $\tilde{\bf{K}}$ which pass the Bonferroni test.
\begin{figure}[ht]
\includegraphics[width=7cm]{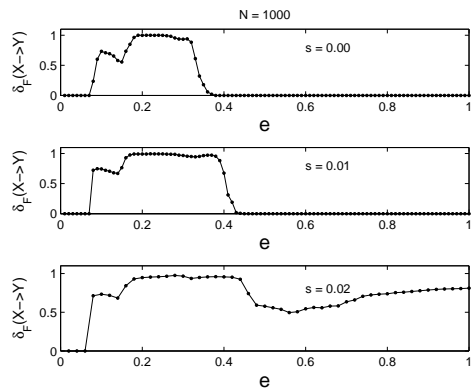}\caption{{\rm
The filtered causality index, for the coupled maps, is displayed
versus $e$ for three values of $s$. The inhomogeneous polynomial
kernel with $p=2$ is used, and $m=1$.
 \label{fig1}}}\end{figure}
We discuss some examples of application of our method with IP
kernel. First we consider two unidirectionally coupled noisy
logistic maps:
\begin{eqnarray}
\begin{array}{l}
x_{n+1}=1-ax_n^2+s \tau_{n},\\
y_{n+1}=(1-e)(1-ay_n^2)+e(1-ax_n^2)+s \gamma_{n};
\end{array}
\label{map}
\end{eqnarray}
$\{\tau\}$ and $\{\gamma\}$ are unit variance Gaussianly distributed
noise terms (the parameter $s$ determines their relevance), $a=1.8$
and $e\in [0,1]$ represents the coupling $x \to y$.  In the
noise-free case ($s=0$), a transition to complete synchronization
\cite{synch} occurs at $e=0.37$. Varying $e$ and $s$, we have
considered runs of $N$ iterations, after a transient of $10^3$, and
evaluated $\delta_F$, using the kernel with $p=2$, in both
directions. We find that $\delta_F\left( Y\to X\right)$ is zero for
all values of $e$, $s$ and $N$. On the other hand $\delta_F\left(
X\to Y\right)$ is zero at $e$ smaller than a threshold $e_c$, see
figure 1. $\delta_F\left( X\to Y\right)$ is zero also at complete
synchronization, as there is no information transfer in this regime.
As noted in \cite{palus}, the causal relation can be inferred only
when the coupling is not large enough to let full synchronization
emerge, or when the synchronized state is frequently perturbed by
internal or external noise driving the system out of the
synchronized state. Indeed, at fixed $e>0.37$, $\delta_F\left( X\to
Y\right)$ is zero until $s$ reaches a threshold $s_c$. Both $e_c$
and $s_c$ scale as $N^{-0.5}$, as expected, see figure 2.
\begin{figure}[ht]
\includegraphics[width=7cm]{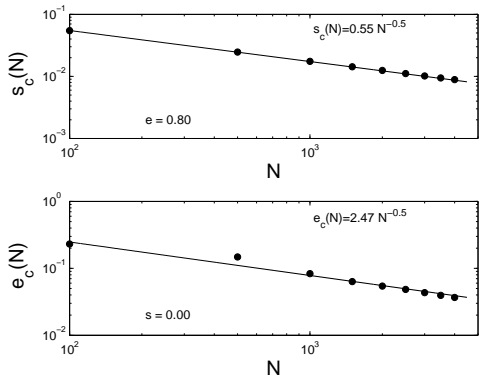}\caption{{\rm
Scaling of the critical values of $e$ and $s$ (see the text) with
$N$. $e_c$ also has a weak dependence on $s$.
 \label{fig2}}}\end{figure}

As another simulated example, we consider two unidirectionally
coupled Henon maps:
\begin{eqnarray}
\begin{array}{l}
   x_n=1.4-x_{n-1}^2+0.3 \hat{x}_{n-1};\hat{x}_n=x_{n-1};\\
    y_n=1.4+0.3 \hat{y}_{n-1}-0.7y_{n-1}^2-0.3 x_{n-1} y_{n-1};\hat{y}_n=y_{n-1};\\
\end{array}
\label{map} \nonumber
\end{eqnarray}
and analyze the causality between  time series $\{x\}$ and $\{y\}$;
by construction, $x$ is driving $y$.  Using $m=2$ and IP kernel with
various values of $p$, on runs of length $N=1000$, we correctly find
that the causality $y\to x$ is always zero whilst the causality
$x\to y$ is non zero and maximal at $p=2$, the interaction being
quadratic (Table I).
\begin{center}
\begin{table}
\begin{tabular}{|c|c|c|c|c|c|}
\hline
p & 1 & 2 & 3 & 4 & 5\\
\hline
$\delta_F$ & 0.04 & 0.88 & 0.81& 0.66& 0.41\\
\hline
\end{tabular}
\caption{Causalities $x\to y$ for coupled Henon's maps.}
\end{table}
\end{center}
A real example consists in rat EEG signals from right (R) and left
(L) cortical intracranial electrodes, employed in the study of the
pathophysiology of epilepsy and already analyzed in \cite{quiroga}.
We analyze both the normal EEG signals and the EEG signals from the
same rat after unilateral lesion in the rostral pole of the
reticular thalamic nucleus, when spike discharges are observed due
to local synchronization of neurons activity in the neighborhood of
the electrode at which the signal was recorded.
\begin{figure}[ht]
\includegraphics[width=7cm]{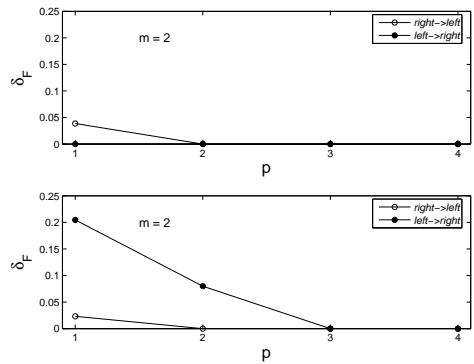}\caption{{\rm
The filtered causality indexes, for the rat EEG signals before (top)
and after the lesion (bottom), is displayed versus $p$, the order of
the inhomogeneous polynomial kernel.
 \label{fig3}}}\end{figure}
In figure 3 top, the indexes $\delta_F$,  for the normal EEG signals
of the rat, are depicted for $p=1,2,3,4$. We find zero causality in
the direction L$\to$R for all $p$, and a small causality R$\to$L
only at $p=1$. After the unilateral lesion (figure 3 bottom) we find
that causality R$\to$L is almost unchanged, whilst a relevant
L$\to$R causality now appears at $p=1$ and (smaller) at $p=2$. A
more conservative statistical procedure, in situations where the
value of $p$ is not known {\it a priori}, is to apply, at each $p$,
the Bonferroni's correction corresponding to the total number of
comparisons, in this case $91$ (=2+9+25+55); using this correction,
the causality R$\to$L becomes zero in both cases and for all $p$,
whilst the causality L$\to$R remains unchanged and equal to the
values depicted in figure 3. The results reported in \cite{quiroga}
are qualitatively consistent with our findings, indeed the same
directions of asymmetry are found in the two analyses, but our
approach allows to make more sharp and precise statements about the
causality relationships between the two EEG signals: the only
statistically robust causality relationship is L$\to$R after the
lesion. Moreover, as the maximum of $\delta_F (L\to R)$ occurs at
$p=1$, our analysis seems to suggest that in this experiment the
information transfer mechanism is essentially linear \cite{nota2}.
\begin{figure}[ht]
\includegraphics[width=7cm]{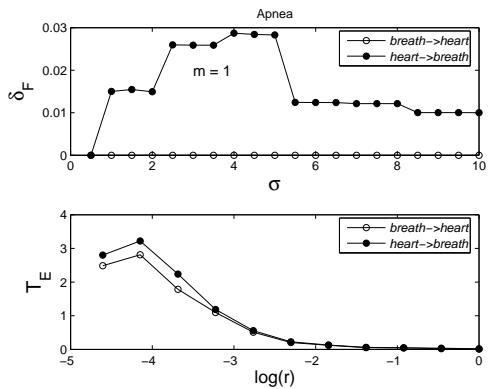}\caption{{\rm (Top)
The filtered causality indexes, for the physionet data-set, are
displayed versus $\sigma$, the width of the Gaussian kernel.
(Bottom) Transfer entropies versus $r$, the length scale.
 \label{fig4}}}\end{figure}

Turning to consider the Gaussian kernel, the condition $H\subseteq
H^\prime$ does not necessarily hold and some differences in the
approach are in order. In this case we call $H$ the
 span of the eigenvectors of $\bf{K}$ whose
eigenvalue is not smaller than $\mu \lambda_{max}$, where
$\lambda_{max}$ is the largest eigenvalue of $\bf{K}$ and $\mu$ is a
small number (we use $\mu= 10^{-6}$). We calculate
$\tilde{\bf{x}}=P{\bf x}$, where $P$ is the projector on $H$. After
evaluating the Gram matrix $\bf{K}^\prime$, the following matrix is
considered: \be {\bf K}^*=\sum_{i=1}^{m_2} \rho_i {\bf w}_i {\bf
w}_i^\top, \label{kstar}\ee where $\{{\bf w}\}$ are the eigenvectors
of $\bf{K}^\prime$, and the sum is over the eigenvalues $\{
\rho_i\}$ not smaller than $\mu$ times the largest eigenvalue of
$\bf{K}^\prime$. Then we evaluate
$\tilde{\bf{K}}=\bf{K}^*-P\bf{K}^*-\bf{K}^*P+P\bf{K}^*P$, and denote
$P^\perp$  the projector onto the range of $\tilde{\bf{K}}$. The
filtered Granger causality index for Gaussian kernels is then
constructed as in the previous cases. As another real example, we
consider time series of heart rate (H) and breath rate (B) of a
sleeping human suffering from sleep apnea (ten minutes from data set
B of the Santa Fe Institute time series contest held in 1991,
available in the Physionet data bank \cite{physionet}). Using IP
kernels, we find unidirectional causality H$\to $B; its strength
increases with the order  $p$ of the kernel, from $\delta_F =0.01$
at $p=1$ to $\delta_F =0.03$ at $p=5$. These findings confirm the
strongly nonlinear nature of the interaction between heart and
respiration signals in sleep apnea syndrome \cite{ja}, which is
evident also using the Gaussian kernel and varying $\sigma$, as
depicted in figure 4 top. No causality B$\to $H is found to be
significative, whilst non zero causality H$\to $B is found for
$\sigma \ge 1$. Note that the causality index vanishes, by
construction, at small $\sigma$ and at large $\sigma$, because in
both limits the kernel matrix tends to be constant ($0$ and $1$,
respectively). In figure 4 bottom the bivariate time series is
analyzed  by means of the transfer entropy \cite{schreiber}. It is
interesting to compare the two approaches in this application. $T_E$
is nonzero in both directions and shows a slightly stronger flow of
information H$\to$B. Our approach recognizes, as significative, only
the causality H$\to$B, thus revealing unidirectional drive-response
relationship in the sleep apnea pathology.

In conclusion, we considered the problem of nonlinear coherence of
signals, in particular the detection of drive-response
relationships. Exploiting the geometry of reproducing kernel Hilbert
spaces we have introduced a {\it filtered} index which is able to
measure cause-effect relationships with arbitrary amount of
nonlinearity, and is not affected by over-fitting.  The choice of
the optimal value of $m$ can be done using the standard {\it cross
validation} scheme \cite{shawe} or the embedding dimension
\cite{kantz}. Our method is equivalent to perform linear Granger
causality in the feature space of the kernel, hence also in the
nonlinear case our approach continues to fulfill the {\it good}
properties of linear models. The framework of Granger causality
assumes stationarity of signals: further work should deal with the
effects of non-stationarities on nonlinear estimates of causalities
(see \cite{ding-prl} for a promising strategy in the linear case).

We expect that the proposed method will provide a statistically
robust basis to assess nonlinear drive-response relationships in
many fields of science, wherever collected data form time series; it
works for deterministic and stochastic systems, provided that noise
is not so high  to obscure the deterministic effects.


\begin{thebibliography}{99}
\bibitem{nota2} Another possible explanation is that, due to high level of
noise, nonlinear effects cannot be put in evidence in this
experiment.
\end{thebibliography}
\end{document}